\documentclass{edp-conf}
\usepackage{graphicx}
\begin{document}
\def\la{\buildrel<\over\sim}
\def\ga{\buildrel>\over\sim}

\TitreGlobal{SF2A 2005}

\title{ACHERNAR CAN BE A DIFFERENTIAL ROTATOR}

\author{Zorec, J.}\address{Institut d'Astrophysique de Paris, UMR7095 CNRS, 
Universit\'e Pierre \& Marie Curie}
\author{Domiciano de Souza, A.}\address{Laboratoire Univ. d'Astroph. de Nice
(LUAN), UMR 6525 CNRS}
\author{Fr\'emat, Y.}\address{Royal Observatory of Belgium}
\author{Vakili, F.$^2$}

\runningtitle{Differential rotation of Achernar}

\index{Zorec, J.}
\index{Domiciano de Souza, A.}
\index{Fr\'emat, Y.}
\index{Vakili, F.}

\maketitle

\begin{abstract}We take advantage of interferometric measurements of Achernar
to inquire on its internal rotational law. The reinterpretation of 
interferome\-tric data and the use of fundamental parameters corrected for 
gravitational darkening effects and models of 2D-models of internal stellar
structures, lead us to the conclusion that the star could not be a rigid, near 
critical, rotator but a differential rotator with the core rotating $\sim3$ 
times faster than the surface.
\end{abstract}

\section{Introduction}

 Achernar ($\alpha$~Eri, HD 10144) is the brightest Be star in the sky and as
such it has deserved detailed observations. Of particular interest are the 
$\lambda$~2.2~$\mu$m interferometric observations carried out by Domiciano de 
Souza et al. (2003), which together with spectroscopic and spectrophotometric 
measurements enable us to inquire on what internal rotational law can account 
for its geometrical deformation.

\section{Method and models}

 Achernar is an active Be star, even during a quiescent or apparent 
non-emission phase. The star had a small H$\alpha$ emission at the epoch of 
interferometric measurements (Vinicius et al. 2005), which imnplies a flux 
excess at $\lambda$~2.2~$\mu$m of rougly 20\% produced by a circumstellar disc.
We may assume, however, that interferometry carried in the stellar polar 
directions are free from disc perturbations. The actual stellar equatorial 
radius can then be estimated requiring apparent area conservation: $R_{\rm 
sph}^2=$ $R_{\rm pole}^{\rm app}(i)\times R_{\rm equat}$, where $R_{\rm sph}$
is the radius of the equivalent circular stellar disc derived by comparing
measured fluxes with model atmospheres. The relation between the apparent 
$i$-dependent interferometric polar radius $R_{\rm pole}^{\rm app}(i)$ and the 
true stellar polar radius $R_{\rm pole}$ is given by:

\begin{equation}
R_{\rm pole}^{\rm app}(i)/R_{\rm equat} = \{1-\bigl[1-(R_{\rm 
pole}/R_{\rm equat})^2]^2\sin^2i\}^{1/2}
\label{rpre}
\end{equation}

\noindent where $i$ is the inclination angle. The use of the $V\!\sin i$ 
parameter corrected for gravitational darkening effects enables us to discuss
the stellar rotational distortion as a function of the internal rotation law
and of the amount of stored angular momentum. Inspired by model predictions of 
rotating stars by Meynet \& Maeder (2000), we adopt the internal rotational 
law given by:
 
\begin{equation}
\Omega(r)/\Omega_{\rm core} = 1-p.e^{-a.r^b} 
\label{rot}
\end{equation}

\noindent where $p$ determines the $\Omega_{\rm core}/\Omega_{\rm surf}$ 
ratio; $b$ determines the steepness of the drop from $\Omega_{\rm core}$ to 
$\Omega_{\rm surf}$; $a$ depends on the size or the core. 2D-models of stellar
interiors were used to determine the changes of the polar and equatorial radii 
as a function of the energy ratio $\tau =$ kinetic energy/$|$gravitational 
potential energy$|$ for a star with mass $M=6.7\pm0.4M_{\odot}$ near the end 
of the MS phase (Vinicius et al. 2005).\par

\section{Results}

 The use of (2.1) and $R_{\rm equat}/R_{\odot}$ against $\tau$ enables
us to compare the observed ($R_{\rm equat}/R_{\odot},R_{\rm equat}/R_{\rm 
pole},V\!\sin i$) with the model predicted ones and determine:

$$\left. \matrix{p &=& 0.624\pm0.001\cr
               \tau&=& 0.014\pm0.001\cr
               \eta&=& 0.69\pm0.07  \cr
          V_{\rm e}&=& 308\pm16 {\rm km/s}\cr
                i  &=& 52^o\pm4^o\cr
\Omega_{\rm core}/\Omega_{\rm surf} &=& 2.7 \cr}\right\} \eqno(3.1)$$ 

\noindent with $\eta =$ $R_{\rm e}^3\Omega_{\rm e}^2/GM$ = ratio of centrifugal
to gravity accelerations. The force ratio is $\eta<1$ which indicates that 
the stellar equator is not at critical rotation. We note that for a plain 
rigid rotation approximation, Achernar would have $\Omega/\Omega_{\rm crit} =$ 
0.8, which corresponds to $\eta_{\rm rigid} =$ 0.66.\par
 It is interesting to note that the energy ratio $\tau =$ 0.014 of our 
differential rotator is quite similar to the value of critical rigid rotators: 
$\tau_c =$ 0.015. In the rigid rotation approximation the star would have 
$\tau_r =$ 0.011, which implies that Achernar is possibly storing 30\% more 
rotational energy that it would have if it were at rigid rotation.\par

\end{document}